\documentclass[twocolumn,final,fleqn]{svjour2}
\usepackage{pslatex}
\usepackage{latexsym}
\usepackage{amsmath,amssymb}
\usepackage{graphicx}
\usepackage{dcolumn}
\usepackage{bm}
\usepackage[numbers]{natbib}
\journalname{Granular Matter}

\begin{document}
\title{Experiments on Corn Pressure in Silo Cells -- 
Translation and Comment of Janssen's Paper from 1895}
\author{Matthias Sperl}
\institute{Duke University, Department of Physics, Box 90305,\\
Durham, NC 27708, USA,\\\email{msperl@duke.edu}
}
\date{\today}

\maketitle 
\begin{abstract}

The German engineer H.A. Janssen gave one of the first accounts of the 
often peculiar behavior of granular material in a paper published in 
German in 1895. From simple experiments with corn he inferred the 
saturation of pressure with height in a granular system. Subsequently, 
Janssen derived the equivalent of the barometric formula for granular 
material from the main assumption that the walls carry part of the weight. 
The following is a translation of this article. The wording is chosen as 
close as possible to the original. While drawings are copied from the 
original, figures displaying data are redone for better readability. The 
translation is complemented by some bibliographical notes and an 
assessment of earlier work, wherein Hagen predicted the saturation of 
pressure with depth in 1852, and Huber-Burnand demonstrated that 
saturation qualitatively as early as in 1829. We conclude with a brief 
discussion of more recent developments resting on Janssen's work.

\end{abstract}

\section{H.A. Janssen, \textit{Engineer in Bremen, Germany}, Experiments 
on Corn Pressure in Silo Cells, 31st August 1895 \cite{janssen95}}

\begin{figure}\begin{center}
\includegraphics[width=.65\columnwidth]{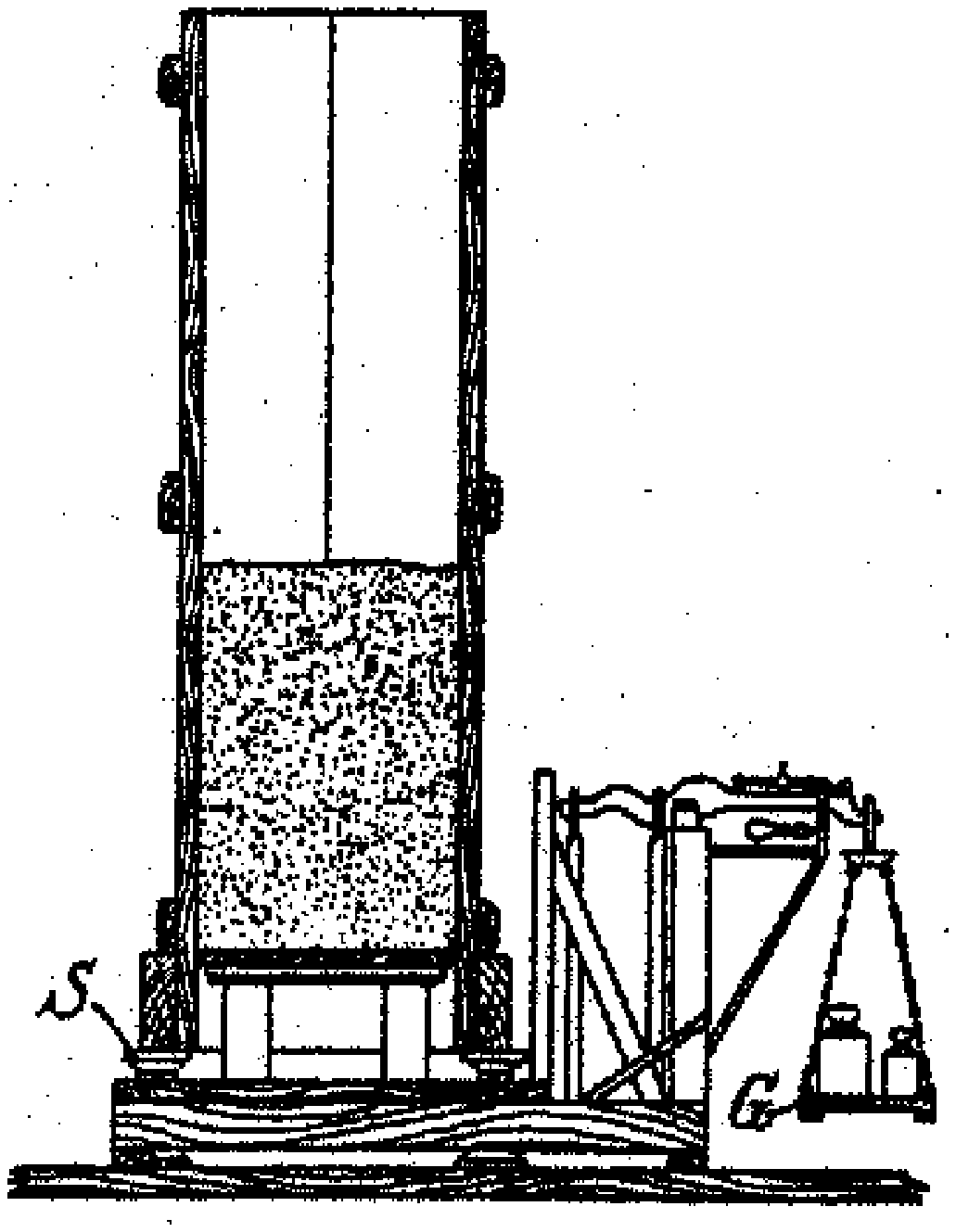}
\caption{\label{fig:F1}}\end{center}
\end{figure}

During the last decades the export of corn from the corn countries of the
world to the culture states of Europe has made extraordinary progress.
Considering the length of the route of transport, this development could
only have been achieved by low production costs, be it due to low wages,
be it due to extensive application of mechanical equipment both for
tillage and for harvesting. Further, consider improvements of corn
depositories and the means of corn transportation. Predominantly in the
United States of North America storage silos have proven themselves to be
beneficial. Constructed on a large scale, these silos are designed to
receive the corn arriving in railway carriages, to store it, and to
dispense it to river and sea vessels \cite{cite1}. The extensive
mechanical equipment for stocking the fruit, weighing it, and loading it
onto ships is highly efficient. Some realizations are capable of handling
several hundreds of tons per hour.  The interior of such a storage silo --
called an \textit{elevator} in North America -- consists of a number of
vertical tubular cavities serving to receive the corn from street-level to
the attic. The profile of these tubes or cells has been exclusively
rectangular with common walls from wooden shafts and bottom casing.
Recently the walls are also produced from iron-strengthened brickwork, and
in six-fold profiles like in honeycombs \cite{cite1}. The grain is
introduced through the upper end of the cell through a hatch. For
discharge, stock transfer, or embarkation, the bottoms of the cells
provide close-able openings. The dimensions of the cells vary
considerably; the largest ones have a capacity of up to around 250~t, in
which case the height of the corn reaches around 25~m. It is obvious that
the enormous content of the cell must exert considerable pressure on the
side walls and the bottom. While various construction guidebooks indicate
the necessity for sufficient anchoring of the cell walls, nowhere is
anything in particular given for the determination of the corn pressure
against the side walls. The formulas valid for the pressure in liquids are
not applicable for corn as the friction among the grains influences the
pressure transmission considerably. Likewise, the formulas customary for
the calculation of the pressure against feed walls are alone for the
reason not applicable for the present case, because the latter are derived
for straight walls, while the content of a silo cell is surrounded by
vertical walls from all sides. To my knowledge, the only publication about
experiments to determine the corn pressure in silo cells were done in the
\textit{Engineering} on October 27th, 1882 \cite{cite2}. These experiments
establish that the pressure of corn in a silo cell against the support
grows with increasing depth up to a certain value, but does not grow
further at larger depth. The largest emerging pressure on the area depends
on the profile of the cell, and is supposedly proportional to the diameter
of the circles inscribed in the profile.

\begin{figure}\begin{center}
\includegraphics[width=.65\columnwidth]{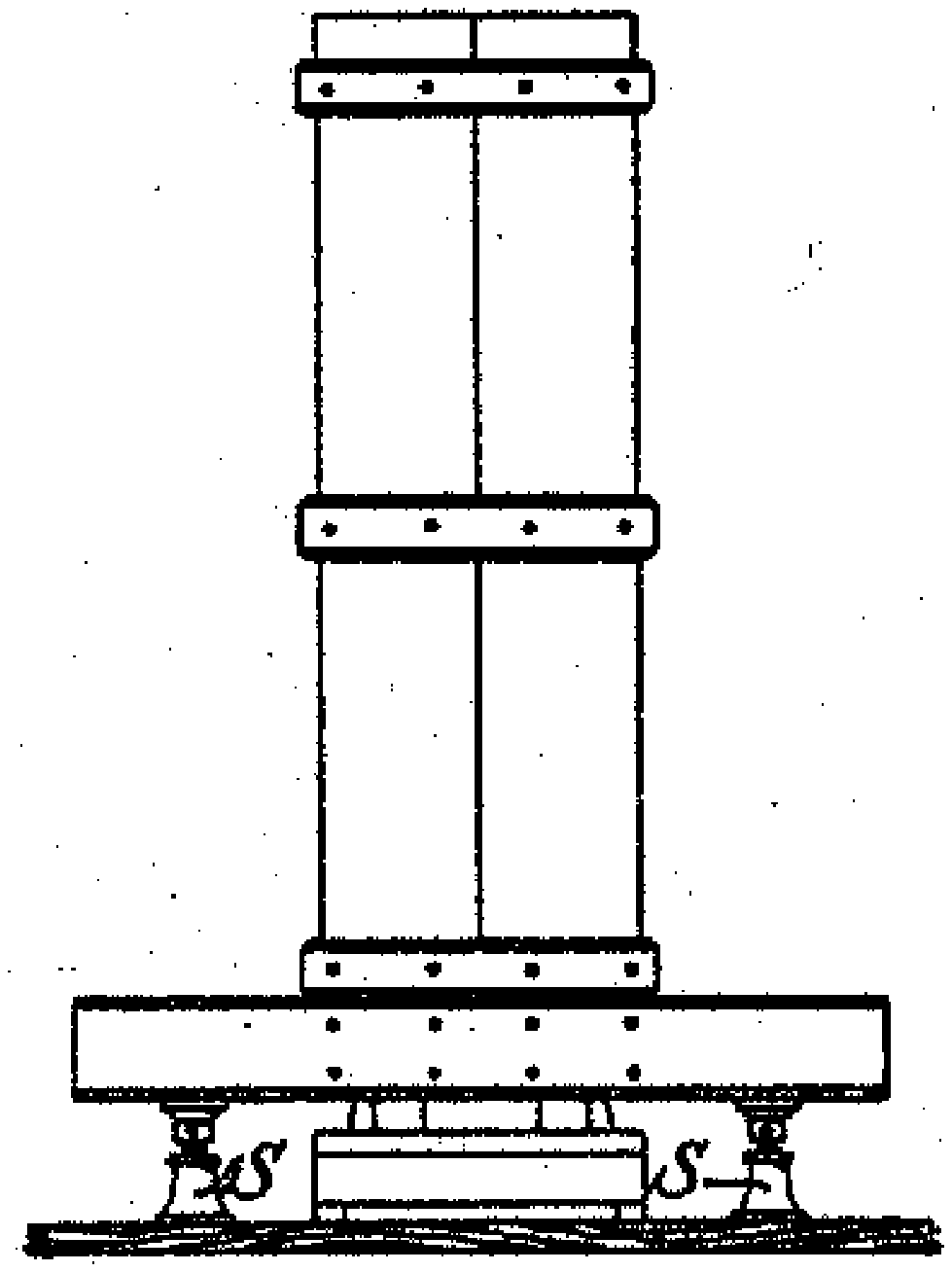}
\caption{\label{fig:F2}}\end{center}
\end{figure}

Since it appeared valuable to me to gain insight about the amount of the
pressure increase in the silo cell from the surface down to arbitrary
depth, I performed detailed experiments, whose results shall be reproduced
here.

\begin{figure}\begin{center}
\includegraphics[width=.35\columnwidth]{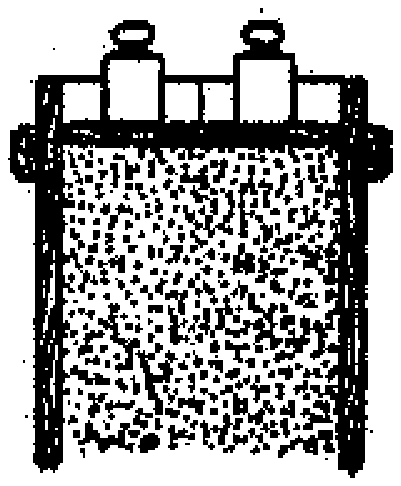}
\caption{\label{fig:F3}}\end{center}
\end{figure}

Four wooden sample cells were produced with quadratic profile and side
lengths of 20, 30, 40, and 60~cm. During the time of the experiment the
respective cell is mounted on four screws $S$, Fig.~\ref{fig:F1} and
~\ref{fig:F2}. The lower end of the tube is secluded by a well-fitting
movable base. This bottom of the cell rests on a decimal balance that is
counterbalanced by weights on the plate $G$ before the experiment. In this
center position a foothold is placed underneath the plate $G$ and $G$ is
loaded with an additional piece of weight. The corn is weighed in hollow
vessels before being poured into the apparatus. Now the cell is filled up
to the point where plate $G$ with the applied weights starts moving
upwards from its center position. Weighing the remaining corn in the
filling container, one can determine accurately the content in the cell
that is causing the now known pressure at the bottom. After placing an
additional weight on the plate $G$, the apparatus is lifted with the
screws $S$ to the point where the balance is in equilibrium again. Now the 
experiment is continued. In this way it was possible to determine the 
bottom pressure for various filling heights while filling the cell only 
once.

\begin{table}\caption{Runs 1--3}
\begin{tabular}{c|c|c|c|c|c}
\hline
\hline
    \multicolumn{2}{c}{run~1, cell~1}	
 & \multicolumn{2}{|c}{run~2, cell~2}
 & \multicolumn{2}{|c}{run~3, cell~2} \\
    \multicolumn{2}{c}{area $20\cdot 20$ cm}
 & \multicolumn{2}{|c}{area $30\cdot 30$ cm}
 & \multicolumn{2}{|c}{area $30\cdot 30$ cm}\\
    \multicolumn{2}{c}{content: wheat}	
 & \multicolumn{2}{|c}{content: wheat} 
 & \multicolumn{2}{|c}{content: wheat}   \\
    \multicolumn{2}{c}{spec. weight $\gamma = 0.8$} 
 & \multicolumn{2}{|c}{spec. weight $\gamma = 0.8$} 
 & \multicolumn{2}{|c}{spec. weight $\gamma = 0.8$}\\
\hline
bottom	& corn	& bottom& corn &bottom& corn	\\
pressure&content&pressure&content&pressure&content\\
kg 	& kg	& kg	& kg	& kg	& kg 	\\
\hline
2.0	& 2.53	& 12.5	& 18.5	& 12.5	& 19.5	\\
2.5	& 3.5	& 14.0	& 22.25	& 15.0	& 27.0	\\
2.7	& 3.9	& 16.5	& 31.8	& 17.5	& 36.5	\\
4.0	& 6.0	& 18.0	& 38.1	& 19.5	& 49.0	\\
4.2	& 7.1 	& 19.0	& 44.4	& 21.0	& 69.7	\\
4.4	& 8.0	& 21.0	& 65.0	& 22.0	& 90.0	\\
4.6	& 9.2	& 22.0	& 78.5	& 23.0	& 180.0	
\footnote[2]{This value is determined like in run~2 after putting weights 
of 22~kg on top of the corn column.}
\\
5.0	& 10.0	& 22.5	& 90.0	&	& 	\\
5.5	& 12.0	& 23.0	& 180.0
\footnote[1]{The apparatus had a capacity of only around 90~kg of wheat,
to be not too unwieldy. After determining the bottom pressure at 90~kg
filling, the surface of the corn was flattened out carefully and a fitting
lid was put on top and loaded with with weights, Fig.~\ref{fig:F3}, in
such a way that the pressure on the surface of the corn was 22.5~kg; that
is the same pressure which is exerted on the bottom for 90~kg filling. The
determined bottom pressure of 23~kg corresponds to a filling of $2\cdot
90$~kg $=180$~kg wheat.}
				&	&\\ 
6.0	& 16.0	& 	&  	&	&\\
6.2	& 17.9	& 	&  	&	&\\
6.4	& 20.0	& 	&  	&	&\\
6.6	& 21.8	& 	&  	&	&\\
6.8	& 24.0	& 	&  	&	&\\
7.0	& 26.0	& 	&  	&	&\\
7.2	& 28.0	& 	&  	&	&\\
7.4	& 36.0	& 	&  	&	&\\
7.5	& 50.0	& 	&  	&	&\\
7.6	& 62.0	& 	&  	&	&\\
\hline
\end{tabular}
\end{table}

\begin{table}\caption{Runs 4--6}
\begin{tabular}{c|c|c|c|c|c}
\hline
\hline
    \multicolumn{2}{c}{run~4, cell~3}	
 & \multicolumn{2}{|c}{run~5, cell~4}
 & \multicolumn{2}{|c}{run~6, cell~4} \\
    \multicolumn{2}{c}{area $40\cdot 40$ cm}
 & \multicolumn{2}{|c}{area $60\cdot 60$ cm}
 & \multicolumn{2}{|c}{area $60\cdot 60$ cm}\\
    \multicolumn{2}{c}{content: wheat}	
 & \multicolumn{2}{|c}{content: wheat} 
 & \multicolumn{2}{|c}{content: wheat}   \\
    \multicolumn{2}{c}{spec. weight $\gamma = 0.8$} 
 & \multicolumn{2}{|c}{spec. weight $\gamma = 0.8$} 
 & \multicolumn{2}{|c}{spec. weight $\gamma = 0.8$}\\
\hline
bottom	& corn	& bottom& corn &bottom& corn	\\
pressure&content&pressure&content&pressure&content\\
kg 	& kg	& kg	& kg	& kg	& kg 	\\
\hline
30	& 40.0	& 80	& 105	& 100	& 140	\\
35	& 52.5	& 120	& 185	& 150	& 290	\\
40	& 66.0	& 140	& 245	& 170	& 430	\\
45	& 85.0	& 155	& 300	& 180	& 500	\\
50	& 106.8	& 165	& 350	& 185	& 540	\\
55	& 138.2	& 170	& 380	& 190	& 890
\footnote[4]{This value is determined like in run~2 after putting weights 
of 165~kg on top of the corn column of 540~kg filling.}	\\
58	& 165.0	& 175	& 440	& 	& 	\\
60	& 192.0	& 180	& 510	& 	& 	\\
63	& 384.0
\footnote[3]{This value is determined like in run~2 after putting weights 
of 60~kg on top of the corn column.}
		& 185	& 540	& 	& 	\\
\hline
\end{tabular}
\end{table}

\begin{figure}\begin{center}
\includegraphics[width=\columnwidth]{F4}
\caption{\label{fig:F4}Run 1.}\end{center}
\end{figure}

\begin{figure}\begin{center}
\includegraphics[width=\columnwidth]{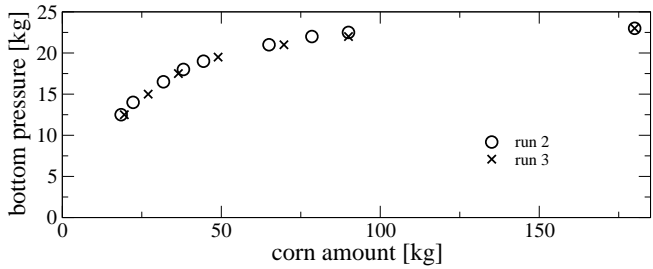}
\caption{\label{fig:F5}Runs 2 and 3.}\end{center}
\end{figure}

\begin{figure}\begin{center}
\includegraphics[width=\columnwidth]{F6}
\caption{\label{fig:F6}Run 4.}\end{center}
\end{figure}

\begin{figure}\begin{center}
\includegraphics[width=\columnwidth]{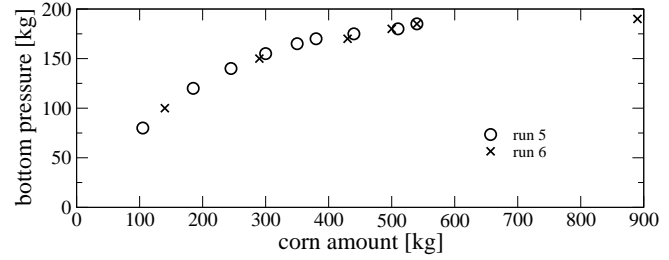}
\caption{\label{fig:F7}Runs 5 and 6.}\end{center}
\end{figure}

The results are displayed graphically in Figs.~\ref{fig:F4} to
\ref{fig:F7}. The smooth path of the observed pressure curve for the runs
2 to 6 lets us conclude that substantial observational errors have not
occurred. Run number~1 -- which was conducted first -- displays relatively
large observational errors, because the observers became more proficient
in the handling of the instruments only during the course of the
experiments.

\begin{figure}\begin{center}
\includegraphics[width=.35\columnwidth]{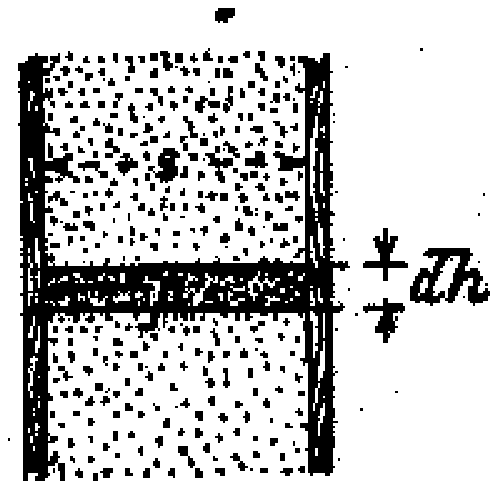}
\caption{\label{fig:F8}}\end{center}
\end{figure}

Were the content of the cell to consist of a liquid, the bottom pressure
would be equal to the weight of the cell's content. The experiments with
corn, however, yield a much smaller bottom pressure. This fact can be
related to the friction between corn and cell walls. This friction becomes
so large at increasing depth that a pressure increase is no longer
noticeable. Hence, the friction between corn and cell walls needs to be
equal to the weight of the enclosed layers of corn. The magnitude of the
corn pressure exerted on the encompassing walls in this case -- the
pressure apparently reaches its highest value -- can be calculated in the
following way. Let $p_{s,\,\text{max}} = $ largest pressure of the corn
against the wall, $f = $ friction coefficient between corn and cell walls,
$s = $ side of the quadratic cell profile, $dh = $ height of a corn layer,
$\gamma = $ specific weight of the corn, then Fig.~\ref{fig:F8} yields the
equation
\begin{equation}\label{eq:eq1}
p_{s,\,\text{max}}\, f\, 4\, s\, dh = \gamma \, s^2\, dh\;,\quad
p_{s,\,\text{max}} = \frac{\gamma \,s}{4 f}\,.
\end{equation}

Assuming a constant coefficient of friction $f$ for variable pressures,
\textit{the maximum pressure exerted on the cell walls by the corn for 
various cell widths of quadratic profile is proportional to the side 
length of the cell profile.} To determine the value of $f$, the following 
experiments were performed:

\begin{figure}\begin{center}
\includegraphics[width=.8\columnwidth]{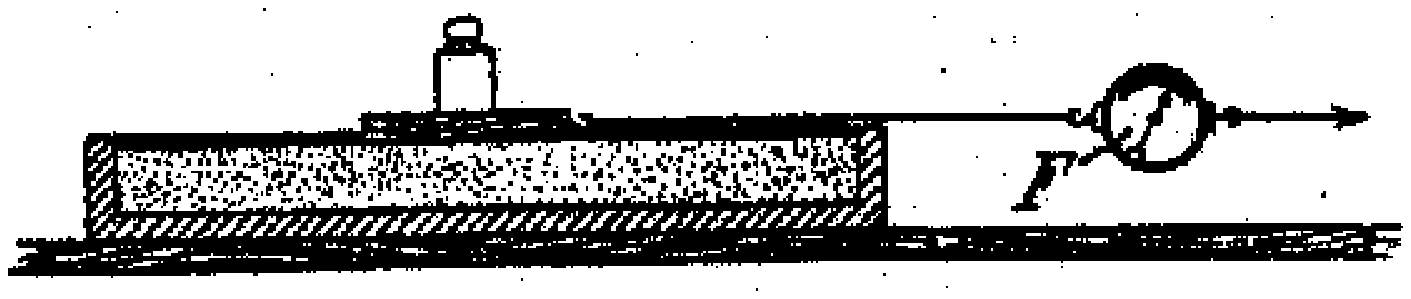}
\caption{\label{fig:F9}}\end{center}
\end{figure}

A flat container, Fig.~\ref{fig:F9}, was filled with wheat and covered
with a wooden plate. The plate could be pulled across the corn with a
string that was connected to a spring balance $F$. By putting weights on
top the pressure between plate and corn could be regulated precisely. The
result of these experiments is given in the following tables.

\begin{table}\caption{Runs 7--9}
\begin{tabular}{c|c|c|c|c|c}
\hline
\hline
    \multicolumn{2}{c}{run~7}	
 & \multicolumn{2}{|c}{run~8}	
 & \multicolumn{2}{|c}{run~9} \\
\hline
pressure&measured&pressure&measured&pressure&measured\\
lid-corn&drag	& lid-corn&drag	& lid-corn&drag	\\
kg 	& kg	& kg	& kg	& kg	& kg 	\\
\hline
5.9	& 1.93	& 10.9	& 3.73	& 20.9	& 6.98	\\
5.9	& 1.93	& 10.9	& 3.73	& 20.9	& 6.73	\\
5.9	& 1.93	& 10.9	& 3.78	& 20.9	& 6.78	\\
5.9	& 1.93	& 10.9	& 3.78	& 20.9	& 7.23	\\
5.9	& 1.98	& 10.9	& 3.83	& 20.9	& 7.28	\\
5.9	& 1.88	& 10.9	& 3.73	& 20.9	& 7.23	\\
5.9	& 2.03	& 10.9	& 3.78	& 20.9	& 7.18	\\
5.9	& 2.08	& 10.9	& 3.78	& 20.9	& 6.98	\\
5.9	& 1.93	& 10.9	& 3.73	& 20.9	& 6.98	\\
5.9	& 1.98	& 10.9	& 3.73	& 20.9	& 6.73	\\
\hline
    \multicolumn{2}{c}{mean drag = 1.96}	
 & \multicolumn{2}{|c}{mean drag = 3.77}	
 & \multicolumn{2}{|c}{mean drag = 7.01}	\\
    \multicolumn{2}{c}{f=1.96/5.9=0.332}	
 & \multicolumn{2}{|c}{f=3.77/10.9=0.346}	
 & \multicolumn{2}{|c}{f=7.01/20.9=0.335}	\\
\end{tabular}\end{table}

\begin{table}\caption{Runs 10--12}
\begin{tabular}{c|c|c|c|c|c}
\hline
\hline
    \multicolumn{2}{c}{run~10}	
 & \multicolumn{2}{|c}{run~11}	
 & \multicolumn{2}{|c}{run~12} \\
\hline
pressure&measured&pressure&measured&pressure&measured\\
lid-corn&drag	& lid-corn&drag	& lid-corn&drag	\\
kg 	& kg	& kg	& kg	& kg	& kg 	\\
\hline
30.9	& 9.23	& 30.9	& 10.22	& 40.9	& 12.53	\\
30.9	& 9.03	& 30.9	& 10.18	& 40.9	& 12.48	\\
30.9	& 9.58	& 30.9	& 10.08	& 40.9	& 12.38	\\
30.9	& 9.23	& 30.9	& 9.93	& 40.9	& 11.93	\\
30.9	& 9.63	& 30.9	& 9.93	& 40.9	& 12.28	\\
30.9	& 9.28	& 30.9	& 10.03	& 40.9	& 12.48	\\
30.9	& 9.43	& 30.9	& 10.38	& 40.9	& 12.28	\\
30.9	& 9.18	& 30.9	& 10.03	& 40.9	& 12.53	\\
\hline
    \multicolumn{2}{c}{mean drag = 9.32}	
 & \multicolumn{2}{|c}{mean drag = 10.10}	
 & \multicolumn{2}{|c}{mean drag = 12.36}	\\
    \multicolumn{2}{c}{f=9.32/30.9=0.302}	
 & \multicolumn{2}{|c}{f=10.10/30.9=0.327}	
 & \multicolumn{2}{|c}{f=12.36/40.9=0.302}	\\
\end{tabular}\end{table}

The investigations No~10 and 11 were done in the same way on different
days. The difference of the friction coefficients determined this way --
there is a deviation of 8\% -- should be ascribed to different humidity.
The maximum value for $f$ resulted from run~8 and was 0.346, the minimum
value from runs 10 and 12 was 0.302.

\begin{figure}\begin{center}
\includegraphics[width=.4\columnwidth]{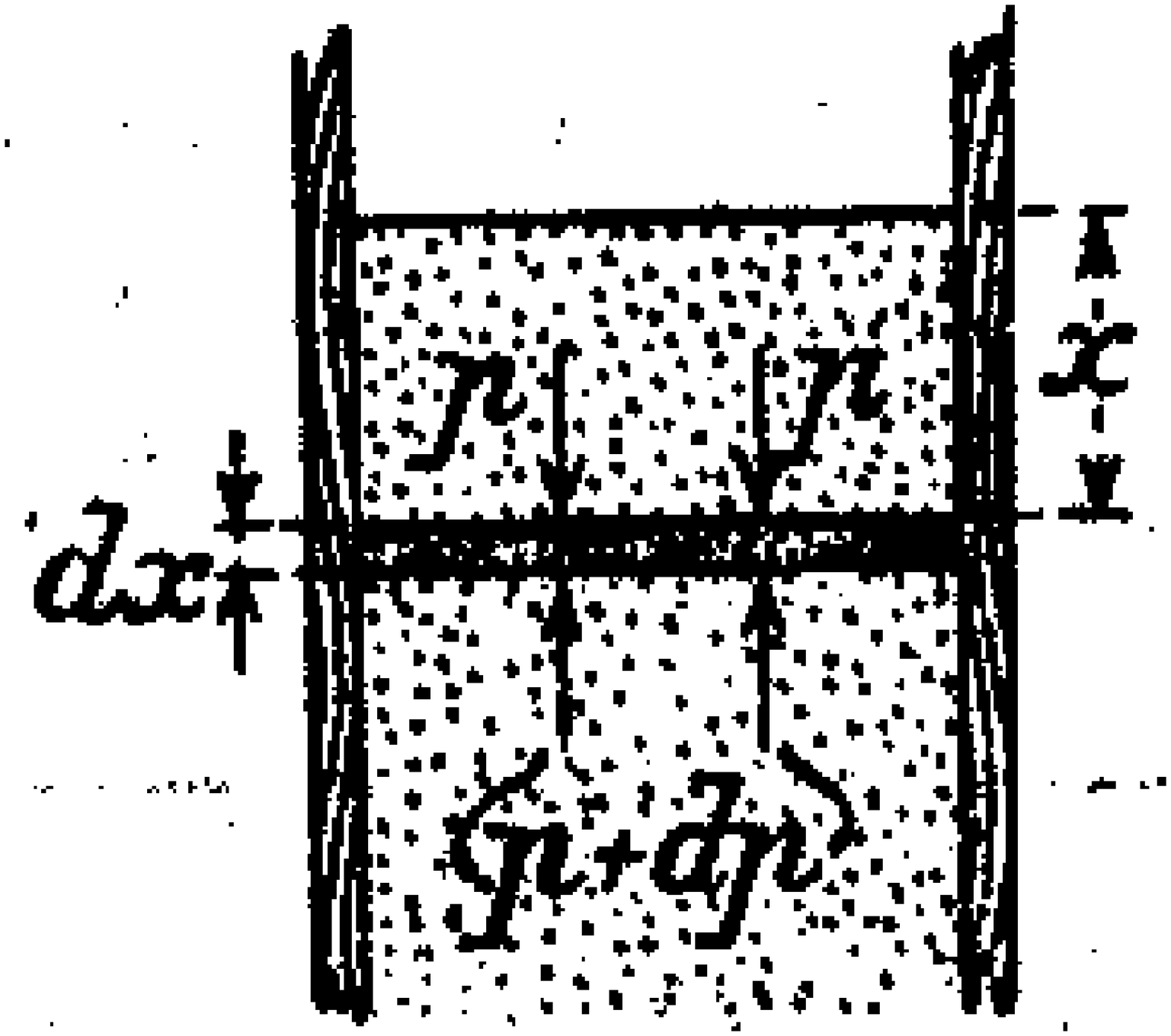}
\caption{\label{fig:F10}}\end{center}
\end{figure}

These investigations where the pressure was gradually increased by a
factor of seven show that the value for the friction coefficient is not
significantly influenced thereby; therefore, $f$ can be assumed constant
with sufficient accuracy for different filling heights. Provided that the
horizontal pressure of the corn is proportional to the vertical pressure,
the latter can be calculated in the following way:

Let $P = $ total pressure of the corn on the cell floor, $p = $ vertical
pressure of the corn, $p_s = $ horizontal pressure of the corn, $f = $
friction coefficient between corn and cell wall, $K = \frac{p_s f}{p}$; $p
K = p_s f$, $s = $ side length of the quadratic cell profile, $u = $
circumference of the cell $ = 4 s$, $F = $ area of the profile $ = s^2$,
$x = $ filling height of the corn in the cell, $\gamma = $ specific weight
of the content, $e = $ base of the natural logarithm; it follows,
Fig.~\ref{fig:F10}:
\[
\begin{split}
F (p + dp - p) = \gamma \, F dx - f p_s u dx\\
dp = \gamma dx -  p \frac{u}{F} dx\\
\frac{dp}{\gamma\left( 1-\frac{K\,u\,p}{F\,\gamma} \right)} = dx\\
-F\ln\left[1-\frac{K u}{F \gamma} p\right] = K u (x-x_0)\\
x=0;\; p=0\\
\ln \left[1-\frac{K u}{F \gamma} p\right] = - \frac{K u x}{F}\\
1 - \frac{K u}{F \gamma}\,p = e^{- \frac{K u x}{F}}\\
1- e^{- \frac{K u x}{F}} =  \frac{K u}{F \gamma} p\\
1- e^{- \frac{4 K x}{s}} =  \frac{4 K}{s \gamma} p
\end{split}
\]
\begin{equation}\label{eq:eq2}
p = \frac{s \gamma}{4 K}\left(1 - e^{-4 K \frac{x}{s}}\right)
\end{equation}
\begin{equation}\label{eq:eq3}
P = \frac{s^3 \gamma}{4 K}\left(1 - e^{-4 K \frac{x}{s}}\right)
\end{equation}

In the latter two equations only $K$ is unknown, which can be determined
from the experiments 1 to 6. In run~4 the largest total bottom pressure
was found at 63~kg or 3.94~kp per 1~qdm. The friction at the circumference
in this case for a layer of 1~dm is $16\gamma = 16\cdot 0.8 = 12.8$~kg
$=p_s f u$;
\[
p_s f = \frac{12.8}{u} = \frac{12.8}{16} = 0.8\;;\quad K = \frac{p_s f}{p} 
= \frac{0.8}{3.94} = 0.203\,.
\]

For simplification I set $K = 0.2$; $\gamma=0.8$. Then Eqs.~(\ref{eq:eq2})
and (\ref{eq:eq3}) assume the following form
\begin{equation}\label{eq:eq2a}\tag{2a}
p = s\left(1 - e^{-0.8 \frac{x}{s}}\right)
\end{equation}
\begin{equation}\label{eq:eq3a}\tag{3a}
P = s^3\left(1 - e^{-0.8 \frac{x}{s}}\right)
\end{equation}

\begin{figure}
\includegraphics[width=\columnwidth]{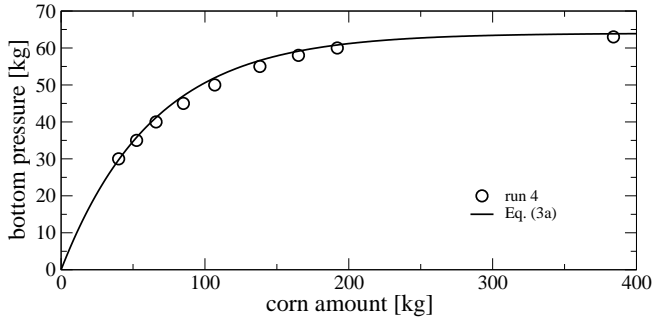}
\caption{\label{fig:F11}Data from run~4 and fit by Eq.~(\ref{eq:eq3a})}
\end{figure}

Figure~\ref{fig:F11} displays the results of run~4 and the calculated corn
pressures according to Eq.~(\ref{eq:eq3a}). Better agreement could have
been achieved by assuming a larger value of $K$, considering the pressure
measurements for smaller filling heights, since it can not be ruled out
that at the observation of the largest pressure an observational error has
occurred. In the same way as described above, $K$ is determined for the
other runs, hence:

\begin{tabular}{ll}
run~1	& $K = 0.211$ \\
run~2, 3& $K = 0.235$ \\
run~4	& $K = 0.203$ \\
run~6	& $K = 0.227$
\end{tabular}

The deviations can be explained by a slight variability of $f$. In runs~2
and~3, $f$ should have been larger by 15.5\% compared to run~4. The
experiments for the determination of the value of the friction coefficient
result in (according to runs~7 to~12) $f=0.302$ to 0.346, or
$14\frac{1}{2}$\% deviation. For the calculation of the side pressure
against the cell wall one can therefore assume that the value of $f$ was
around 0.346 in run~2 and~3 and around 0.302 in run~4. One gets from $p_s
= \frac{K p}{f}$ for run~2 and~3: $P_s = \frac{0.235}{0.346}p = 0.68 p$,
and for run~4:  $p_s = \frac{0.203}{0.302} p = 0.675 p$, or rounded $p_s =
0.7 p$

\begin{figure}
\includegraphics[width=\columnwidth]{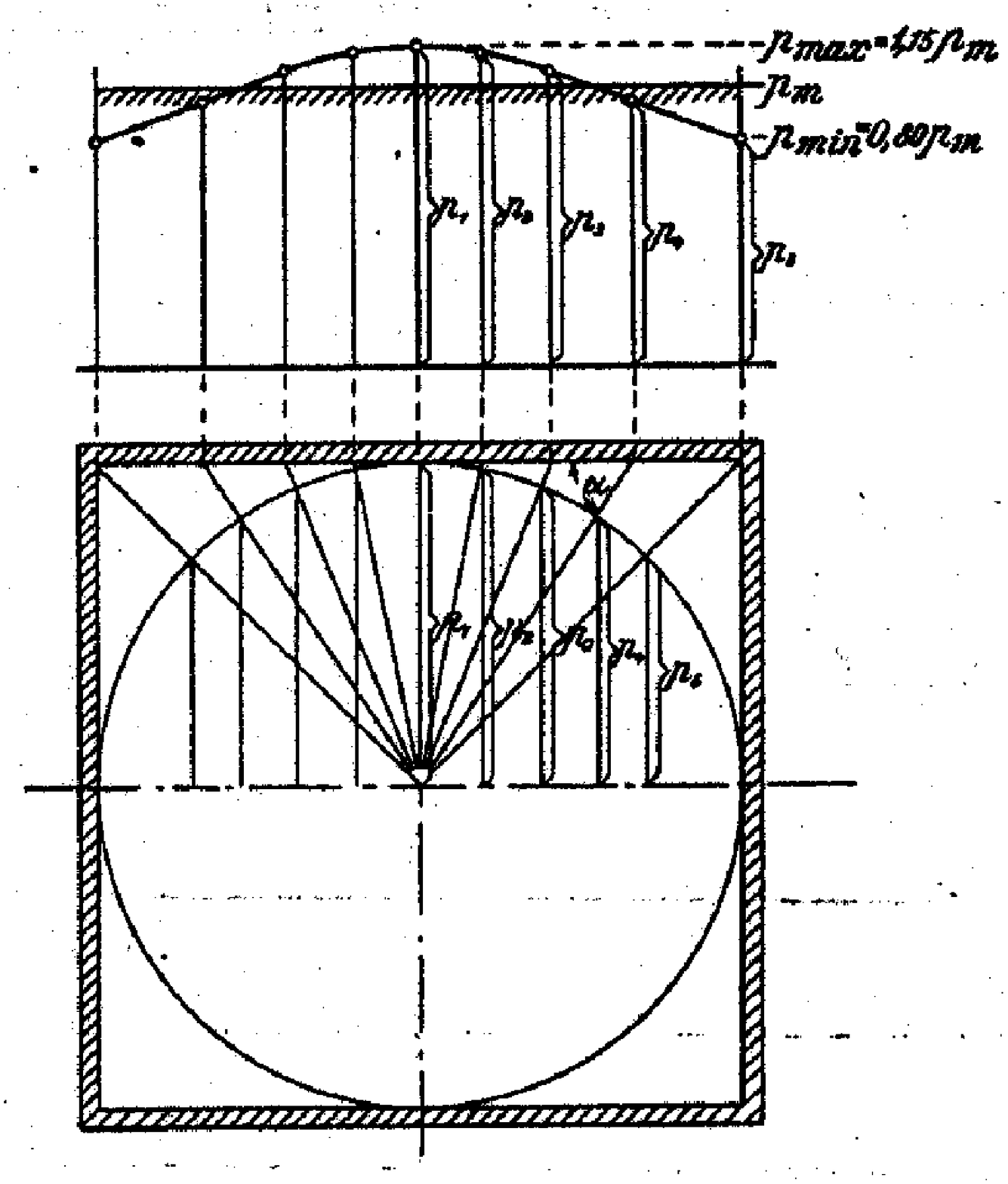}
\caption{\label{fig:F12}}
\end{figure}

Hereby the \textit{mean} pressure against a side wall of a cell of
quadratic profile is determined.  However, we can safely assume that near
the corners of the profile the pressure is lower than this mean value,
while it is higher in the center part of the walls. It is shown in
Fig.~\ref{fig:F12} how the pressure is transmitted to the side walls.
Thereby it was assumed that the distribution of the corn pressure
originates radially from the center of the cell and exerts a pressure
against the wall of $p\sin\alpha$, with $\alpha$ being the angle under
which the pressure ray hits the wall. The maximum pressure against the
wall is reached in the middle of the cell wall with around 1.15 of the
mean pressure, or $=1.15 \cdot 0.7 \cdot p \approx 0.8p$. For the
calculation of the wall thickness the ansatz of \textit{uniform loading}
can be made with sufficient accuracy
\begin{equation}\label{eq:eq4a}\tag{4a}
p_s = 0.75 p = s\cdot 0.75 \left( 1 - e^{-0.8\frac{x}{s}} \right)
\,.
\end{equation}

\begin{figure}
\includegraphics[width=\columnwidth]{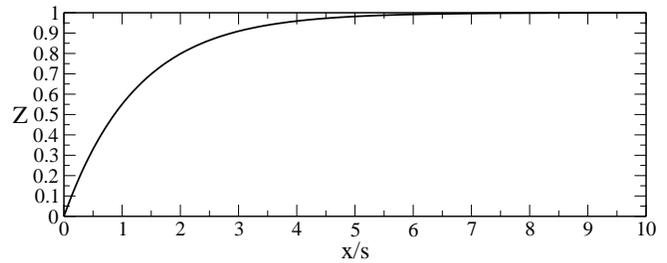}
\caption{\label{fig:F13}Calculation of the bottom pressure.}
\end{figure}

Even if the above experimental and calculated results are sufficient to
examine silo bottoms and silo walls for their necessary stiffness, a
simpler way of calculation might be desirable. We are on the safe side for
the usual kinds of corn if we assume wheat as cell content with a specific
weight of 0.80, set $K = 0.20$, and calculate the corn pressure at the
bottom according to formulas~(\ref{eq:eq2a}) and~(\ref{eq:eq3a}). From the
graph in Fig.~\ref{fig:F13} one can obtain the value for the brackets in
Eqs.~(\ref{eq:eq2a}) and~(\ref{eq:eq3a}) $ = \left( 1 -
e^{-0.8\frac{x}{s}} \right) = Z$ for arbitrary $\frac{x}{s}$.

\begin{figure}
\includegraphics[width=\columnwidth]{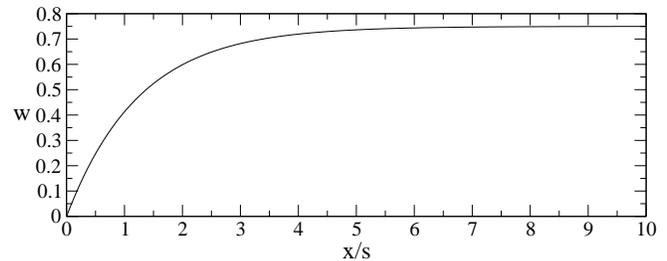}
\caption{\label{fig:F14}Calculation of the side pressures}
\end{figure}

From the just determined vertical pressures one could get the respective
side pressures -- which are decisive for the dimensioning of the cell
walls -- by multiplying by 0.75; for greater convenience the value of $w =
0.75 \left( 1 - e^{-0.8\frac{x}{s}} \right)$ can be obtained from
Fig.~\ref{fig:F14} for arbitrary $\frac{x}{s}$.

\begin{table}
\caption{\label{tab:ex1}
Example~1. Calculation of the bottom pressure for a cell of $4\times 4$~m
base: $s=4$. $Z$ from Fig.~\ref{fig:F13}, $p=sZ$, $P=s^3Z$,
}
\begin{tabular}{@{\extracolsep{5mm}}rlccc}
filling 
$x/$m	& $x/s$	& $Z$	& $p$	& $P$ \\
\hline
1	& 0.25	& 0.18	& 0.72	& 11.5 \\
2	& 0.50	& 0.33	& 1.32	& 21.1	\\
3	& 0.75	& 0.45	& 1.80	& 28.8	\\
4	& 1.00	& 0.55	& 2.20	& 35.2	\\
5	& 1.25	& 0.63	& 2.52	& 40.3	\\
6	& 1.50	& 0.70	& 2.80	& 44.7	\\
8	& 2.0	& 0.80	& 3.20	& 51.2	\\
10	& 2.5	& 0.86	& 3.44	& 55.0	\\
12	& 3.0	& 0.91	& 3.64	& 58.2	\\
14	& 3.5	& 0.94	& 3.76	& 60.1	
\end{tabular}
\end{table}

\begin{table}\caption{\label{tab:ex2}
Example~2. Calculation of the side pressure of a silo cell of $3\times 
3$~m base: $s=3$. $w$ after Fig.~\ref{fig:F14}, $p_s=s w$,}
\begin{tabular}{@{\extracolsep{5mm}}rrrl}
filling 
$x$[m]	& $x/s$	& $w$	& $p_s$ [t/m$^2$] \\
\hline
1.5	& 0.50	& 0.245	& 0.735	\\
3.0	& 1.00	& 0.410	& 1.23	\\
4.5	& 1.50	& 0.530	& 1.59	\\
6.0	& 2.00	& 0.595	& 1.79	\\
8.0	& 2.67	& 0.655	& 1.97	\\
10.0	& 3.33	& 0.690	& 2.07	\\
12.0	& 4.00	& 0.720	& 2.16	
\end{tabular}
\end{table}

Originally it was the intention of the author to determine the side
pressure of the corn directly in the experiments. To this end, one of the
sample cells was equipped with a side lid that was pressed on the cell by
suitable weights by an angle lever. The pressure necessary to open the lid
could then be calculated. But while conducting the experiment the opening
of the lid happened so slowly that accurate results could not be obtained.
The same occurred with the attempt to measure the bottom pressure at
various points on the bottom via small lids. In my opinion this feature
can be traced back to the fact that arching occurs as the profile of the 
corn column narrows, which significantly influences the pressure 
transmission for smaller profiles. The attempt to equip the entire 
experiment from top to bottom with a movable cell wall was discarded 
because substantial inconveniences would have been incurred.  Also the 
calculation method described above led to the same goal.

Besides wheat, further investigations were conducted with rye and maize,
whose results are in agreement with the experiments described above. With
rye -- with a specific weight of 0.75 -- the pressures fell short of the
ones for wheat by around 20\%. Maize at the same specific weight as wheat
($\gamma = 0.80$) produced a larger bottom pressure by 22\% due to its
smoother grain surface. For a silo cell that is supposed to contain maize,
the stiffness of the cell walls and the bottom must be increased by 22\%.

The experiments described above -- from which the calculations for
arbitrary cell sizes are deduced -- are performed in small apparatuses
only, because the creation of larger sample cells is connected with
non-negligible costs. Regardless, they should be clarifying in some
aspect, and it is hoped that they find confirmation in experiments on
larger scale like the ones announced by a well-known mill-building firm. I
shall follow up with a further communication on that subject at that
point.

\section{Janssen and Predecessors}

Janssen cites only two earlier works, and these are given rather 
incompletely. However, they are believed to be cited correctly here. The 
first is Ref.~\cite{cite1}, \textit{Die Silospeicher von Galatz und 
Braila} -- \textit{The silos of Gala\c{t}i and Br\u{a}ila}, where 
Gala\c{t}i and Br\u{a}ila are two Romanian cities at the Danube river. The 
second reference is \cite{cite2}, \textit{The Pressure of Stored Grain}, 
and can be seen as an earlier account of the saturation effect observed in 
these silo experiments. This work is clearly acknowledged by Janssen, so 
it may seem unjustified to call this saturation effect alone the 
\textit{Janssen effect}.

There exists an earlier partial translation of Janssen's paper in 
\cite{trans0}, \textit{On the Pressure of Grain in Silos}, which is listed 
there under the category of \textit{Foreign Abstracts} and contains a 
short summary of the article by an editor. According to this editor, the 
saturation of the pressure "[\dots] has long been recognised". While 
Janssen's experiments are described in somewhat more detail, the 
Eqs.~(\ref{eq:eq2}) and (\ref{eq:eq3}) are only cited without derivation 
and are indeed copied wrongly from the original.

In fact, one can infer from earlier work that is not cited by Janssen, 
that the saturation of the pressure with depth was already more or less 
well known in Janssen's time. Janssen's work is anticipated in 1852 by a 
paper of G.H.L.~Hagen \cite{Hagen1852}, who became famous for his work in 
fluid dynamics, and an even earlier investigation by Huber-Burnand in 1829 
\cite{Huber1829}. Hagen imagined a cylinder of radius $r$ and height $h$ 
inside a container filled with sand of a specific weight $\gamma$. The 
pressure exerted on the bottom area $\pi r^2$ by the weight of the 
sand-cylinder is diminished by the friction experienced by that cylinder 
from the sand surrounding it. Assuming isotropic horizontal pressure 
throughout the system, Hagen derives the formula
\begin{equation}\label{eq:hagen}
p(h) \propto r^2\pi\gamma h - 2 r \pi\gamma l h^2
\end{equation}
with some friction coefficient $l$. He discards the decreasing branch of 
the solution and predicts the saturation of the pressure with depth after 
the height reaches the value $h=r/(4l)$. Using the latter relation, Hagen 
subsequently tests his predictions in an apparatus quite similar to 
Janssen's setup. He determines values for $l$ between 0.154 and 0.22, but 
he does not check if the pressure below the maximum follows 
Eq.~(\ref{eq:hagen}) which predicts a steeper increase than Janssen's 
Eq.~(\ref{eq:eq2}). Hagen rather continues to analyze in detail the flow 
through an opening of his container and finds that the mass flow is 
proportional to some effective radius to the power 5/2, cf. 
Ref.~\cite[sec.~10.2]{nedderman92}. For both flow and pressure 
measurements, Hagen mentions earlier work by Huber-Burnand without giving 
any reference. The appropriate reference is most likely 
Ref.~\cite{Huber1829} which discusses largely qualitative results of 
experiments with sand, peas, and -- an egg. The egg is placed in a box, 
covered with several inches of sand, and loaded with a weight of 25~kg; 
the egg demonstrates -- by not breaking, of course -- that only a small 
fraction of the pressure from the top actually reaches the bottom.

\section{Impact of Janssen's paper}

\begin{figure}
\includegraphics[width=\columnwidth]{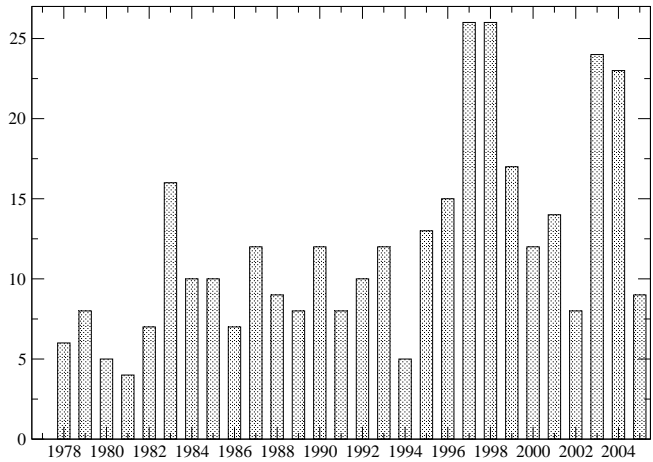}
\caption{\label{fig:cit}Citations of the paper \cite{janssen95} since 
1978 \cite{isi}.
}
\end{figure}

The journal where Janssen published his results can still be found today 
\cite{vdi} -- it is still a publication of the \textit{Verein Deutscher 
Ingenieure} -- the \textit{Association of German Engineers}. 
Figure~\ref{fig:cit} shows the citations of Janssen's paper 
\cite{janssen95} in the years since 1978. As of September~1st~2005, the 
article was cited 375 times with only 40 citations recorded for the time 
1977 and earlier \cite{isi}. This demonstrates growing interest in 
granular systems in general and in one of its earliest papers in 
particular.

It is interesting to note that Janssen's formula~(\ref{eq:eq2}) is usually 
derived for the more symmetric cylindrical geometry 
\cite[sec.~5.2]{nedderman92} and that more involved profiles require some 
averaging of the wall stress along the perimeter, which can be done 
explicitly, cf. \cite[sec.~5.5]{nedderman92}, or implicitly as in 
\cite[sec.~3.1.4]{Duran2000}. In any case, the profile does only change 
the prefactors in Eq.~(\ref{eq:eq2}) but the overall behavior remains the 
same. Since he was experimenting with a square profile, Janssen was well 
aware of that kind of complication as seen in his more involved argument 
connected to Fig.~\ref{fig:F12}. It was also found empirically that the 
saturation pressure was proportional to the diameter of the circle 
inscribed into the profile by Roberts \cite{cite2}. Janssen reports this 
earlier finding, and Fig.~\ref{fig:F12} may suggest that he considered it 
essentially correct.

Janssen's model rests on a number of approximations that were scrutinized 
later on:
First, Eq.~(\ref{eq:eq1}) amounts to full mobilization of friction at the 
walls. While that is not true in general, it might well hold to a good 
degree for Janssen's experimental protocol where the walls of the 
container were shifted upwards by the screws $S$ in Figs.~\ref{fig:F1} and 
\ref{fig:F2} after each step before filling in more grain.
Second, the vertical and horizontal stresses are introduced as principal 
stresses which is clearly inconsistent with the existence of a finite 
friction at the walls. These principal stresses -- horizontal and vertical 
pressures $p_s$ and $p$, respectively -- are then related by a 
constitutive law, $p K = p_s f$, which is not motivated any further.
Third, a uniform distribution of stresses is assumed for any slice of the 
silo, cf. Fig.~\ref{fig:F10}. Taking the constitutive law for granted, 
this translates into a finite shear stress in the center of the silo which 
must be ruled out by symmetry.
All three of these objections can be overcome by more refined continuum 
theories \cite{nedderman92}, and the result in Eq.~(\ref{eq:eq2}) remains
valid with appropriate changes to the prefactors. There are, however, more 
serious problems when considering relatively large overloads on top of the 
grains in the silo \cite[sec.~5.5]{nedderman92}. 

If one probes in more detail the granular structure within the silo and 
especially when considering the dynamics of the granular material, 
Janssen's approach no longer yields satisfactory answers \cite{Duran2000}. 
That said, it is quite remarkable that Janssen already postulates that it 
is arching within the granular assembly what prevented him from measuring 
the bottom pressure at various points.

\begin{acknowledgements}
I want to thank B.P.~Tighe for careful proofreading of the translation, 
Springer-VDI~Verlag for granting the permission to publish this translation,
S. Luding and Springer-VDI~Verlag for assistance with the references, and 
the referee for pointing out Ref.~\cite{Hagen1852} to me.
This work is supported by NSF-DMT0137119, DMS0244492, and DFG~SP~714/3-1.
\end{acknowledgements}


\end{document}